# Disparities in greenspace access during COVID-19 mobility restrictions

David Lusseau[1] & Rosie Baillie[2]

[1]National Institute for Aquatic Resources, Technical University of Denmark, Kgs. Lyngby 2800, Denmark

[2]School of Biological Sciences, University of Aberdeen, Aberdeen AB24 2 TZ, UK

## Abstract

More than half of the human population lives in cities[1] meaning that most people predominantly experience nature in urban greenspace. Nature exposure is an important contributor to social, mental and physical health[2–9]. As the world faces a pandemic which threatens the physical and mental health[10] of billions of people, it is crucial to understand that all have the possibility to access nature exposure to alleviate some of these challenges. Here, for the first time, we integrate data from Facebook, Twitter, and Google Search users to show that people looked for greenspace during COVID-19 mobility restrictions but may not have always managed to reach it. People spent more time in areas with greenspace when they could and that depended on the level of multiple deprivation in the neighbourhood where the greenspace was embedded. Importantly, while people sought greenspace throughout the first 20 months of the pandemic, this preference intensified through the waves of lockdown. Living in an affluent area conferred a greenspace advantage in London and Paris but we find that in Berlin more deprived neighbourhoods sought greenspace more, including outside their neighbourhood. This highlights the need to understand how greenspace access and deprivation interact to create more sustainable communities.

Over the past decade the link between access to some vegetated space or waterbodies in the urban landscape (greenspace) and both physical and mental health has become apparent[2,11–15]. Studies have been able to disentangle greenspace exposure from physical activity confounding effects to show that exposure to features of wild and urban nature-rich spaces has an effect on physical, mental and social health[2–9]. These features include both habitats and species[11,16–18] which coproduce these benefits with a wide range of human activities from sports to simple contemplation[19]. There are however inequalities in greenspace access and quality associated with deprivation[20,21] and urban plan features[22] of neighbourhoods. Also, people can only access greenspace when they have free time coinciding with when the space is freely and safely accessible[23–25].

More than half the human population now lives in cities[1] and therefore has regular access to nature and its cultural ecosystem services[26] only in these anthropogenic landscapes[6,16,27]. It is therefore crucial to understand how people use greenspace to maximize the equitable provision of the health benefits it confers. Indeed, this is a key 2030 target for the United Nations Global Goal (SDG) 11. While we can estimate in many cities where greenspace is, we generally lack an understanding of its accessibility and access[22].

Survey-based studies have shown that people have tended to seek greenspace more during the mobility restrictions associated with the first wave of the SARS-CoV-2 pandemic (COVID-19) in several countries[28–32]. However, there were inequalities in the ability to realise this increased interest in greenspace, and some people spent less time in greenspace despite understanding its health importance[23].

This counterfactual show that realised greenspace use is complex and equitable access is likely affected by deprivation factors. COVID-19 offered, unfortunately, a unique natural experiment to assess the interest of people for greenspace and their ability to meet this need throughout the multiple waves of the pandemic. Here we used the unique Facebook Population data from the Data for Good programme[33] to assess whether people used greenspace more over the pandemic period depending on mobility restrictions (Extended Data Fig. 1) in three European capital cities: London, Paris, and Berlin. Those cities were selected because both indices of multiple deprivation (IMDs)[34] and curated public greenspace data were openly available (see Methods) at a spatial resolution relevant to understand the socioecological context of human mobility[35].

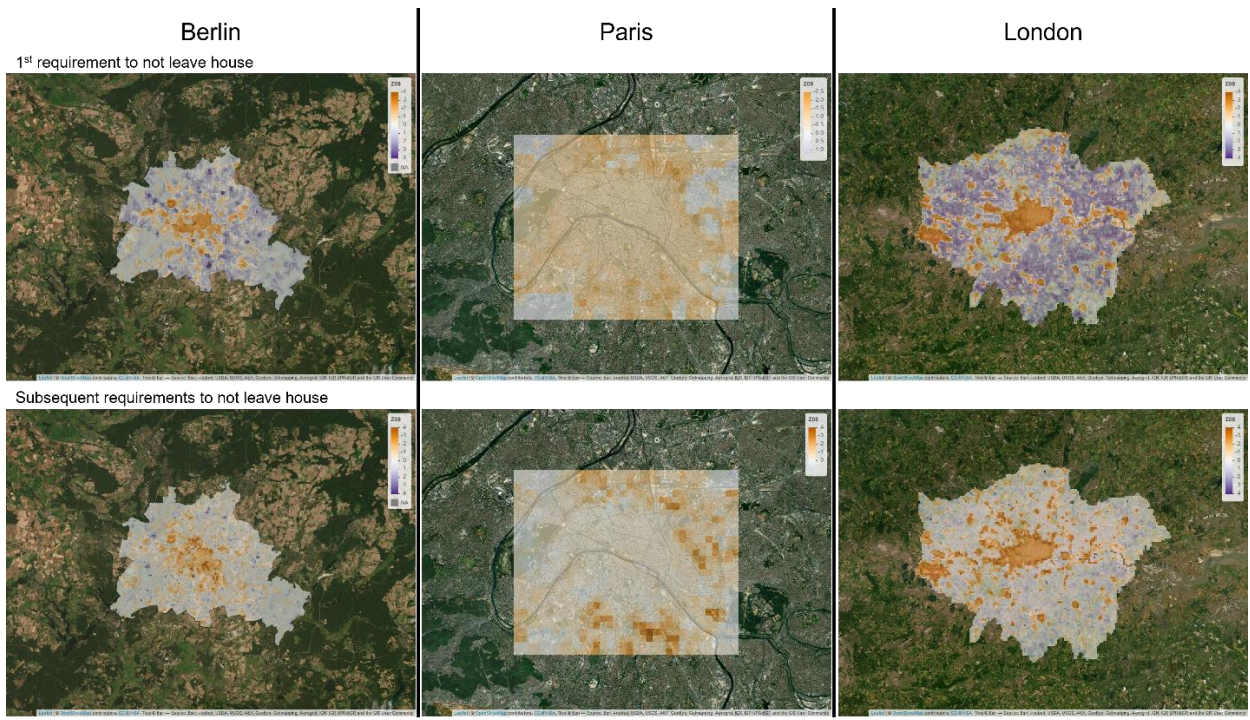

Figure 1. Changes in Facebook population density in Berlin, Paris and London during the day (Z-score, $Z_{08}$) compared to pre-pandemic levels during the first period when people were required not to leave home (March-May 2020 'lockdown') and during subsequent lockdowns over the study period (post September 2020 'lockdowns', see Extended Data Figure S1 for details).

The Facebook population data aggregates the number of the app users that have their location history enabled to 'tiles', quadrat of specific size of which the resolution depends on population density to ensure user anonymity[33]. In large cities this corresponds to quadrat with sides of about 300-500m (depending on latitude), a relevant scale for human urban mobility[35]. We could also determine for each tile, the proportion of greenspace and the median index of multiple deprivation. The Facebook population is sampled in three 8-hour bins. Here we used the Facebook Z-score value to estimate how different from 'normal conditions', that is pre-pandemic times, the population density was on a particular day and for a particular time bin[33] (Fig. 1). This gave us a direct measure of whether a tile was used more than "normal" during the pandemic. There could be an association between the Z-score and the proportion of greenspace in a tile simply due to people being restricted to their homes and residential areas potentially having more greenspace. We therefore run two sets of models, one looking at the variation in the Z-score of tile for the daytime period 08:00-16:00 ($Z_{08}$) and one contrasting this Z-score to the Z score for the night-time period 00:00-08:00 ($Z_{08-00}$), time at which we are sure that most people were most likely to be at home. If the greenspace effect was only caused by the increased propensity of staying at home during lockdowns, then we should not see a greenspace effect on this latter measure.

**Urban mobility and greenspace**

Given the health benefits of greenspace exposure, we expected that the richer in greenspace a tile was, the more people sought it during all mobility restriction periods, but that this effect would depend on the time people had available for leisure[36]. Hence, we would expect increased use during weekends everywhere, and at all times in affluent areas. We first found that greenspace rich tiles, associated with the main parks and gardens in the cities, were similarly used more during the first lockdown than in subsequent ones except in Paris where parks and gardens were closed by decree during the first wave[37] (Extended Data Fig 2).

We found that for all three cities, variance in $Z_{08}$ and $Z_{08-00}$ could be best explained by the same model which considered the effect of lockdown measures to depend on greenspace coverage, IMDs, and whether it occurred on a weekend (Extended Data Figs. 3-8). However, there was substantial variance in all three cities associated with lockdown waves (Fig. 2). The patterns observed during the first lockdown did not repeat in subsequent ones. Through time, tile density decreased in all cities. The pattern of lockdown waves differed between the cities. In London, the preference for greenspace in affluent areas observed in the first lockdown waned in subsequent waves while the preference observed in deprived areas in Berlin increased, particularly in weekends (Fig. 2). There was a tendency for preference for tiles depending on their greenspace coverage to increase with tile affluence in Paris, but that effect was disparate between the COVID-19 waves. This difference further reinforces the observed associations being driven by greenspace access as Paris closed access to its greenspace during the first wave[37].

Berlin has large forest parks in its district which are not curated as greenspace because they are managed by a federal agency. It may therefore be that individuals from affluent neighbourhoods used these forests more than greenspace in their neighbourhood. Using Facebook's movement data (see Methods) we could assess from which tile people visiting forests were and therefore from which neighbourhood level of deprivation. We found that this was not the case. Forests were used significantly more than usual by people originating from deprived areas (Extended Data Fig. 9), particularly over weekends. Forests were visited significantly less than usual by people from affluent areas during weekdays. Berlin has neighbourhood-embedded open access community gardens[38]. It may be that this unique feature helped explain the greater propensity for public access to greenspace in more deprived areas in this city during the lockdowns.

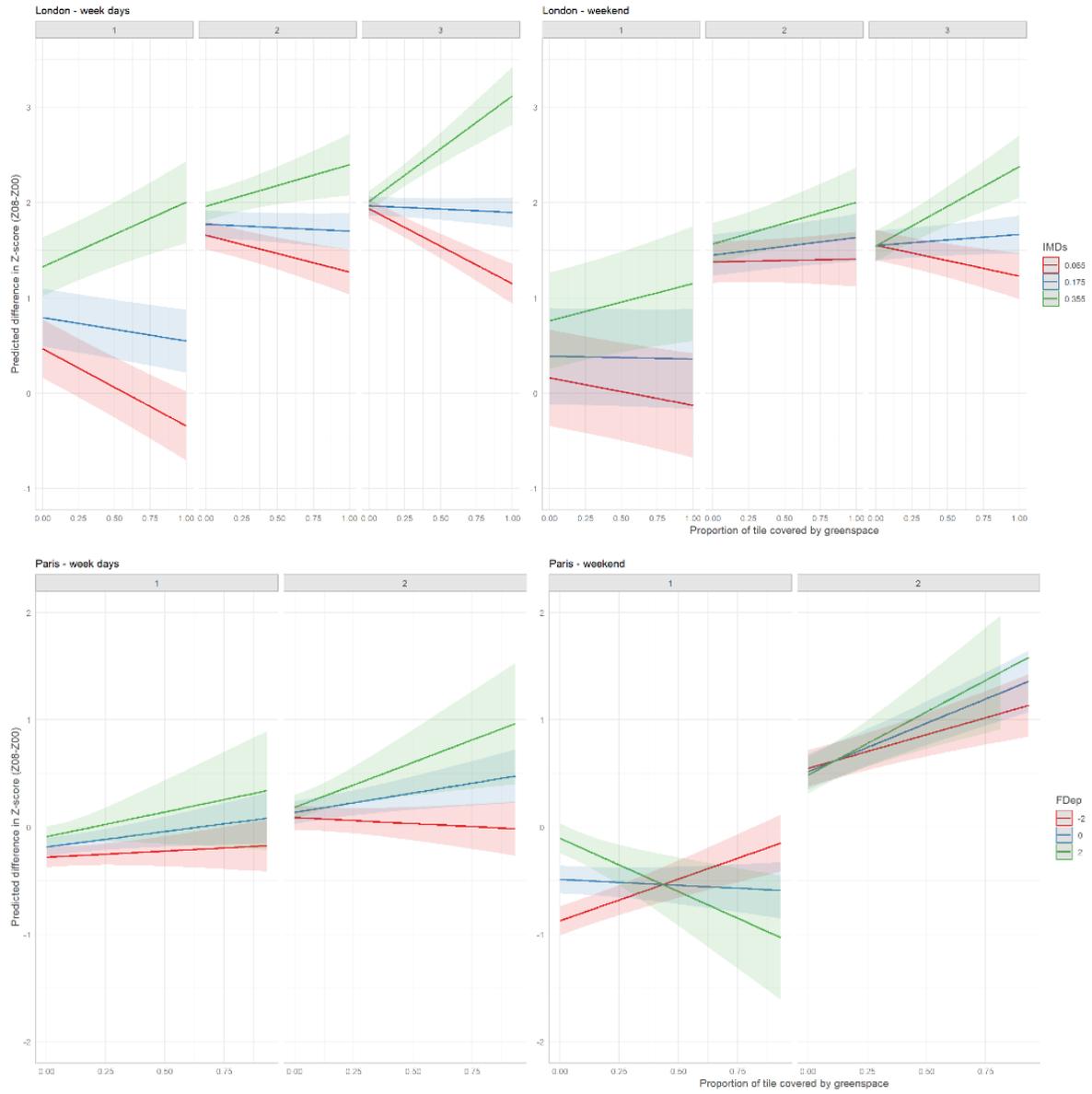

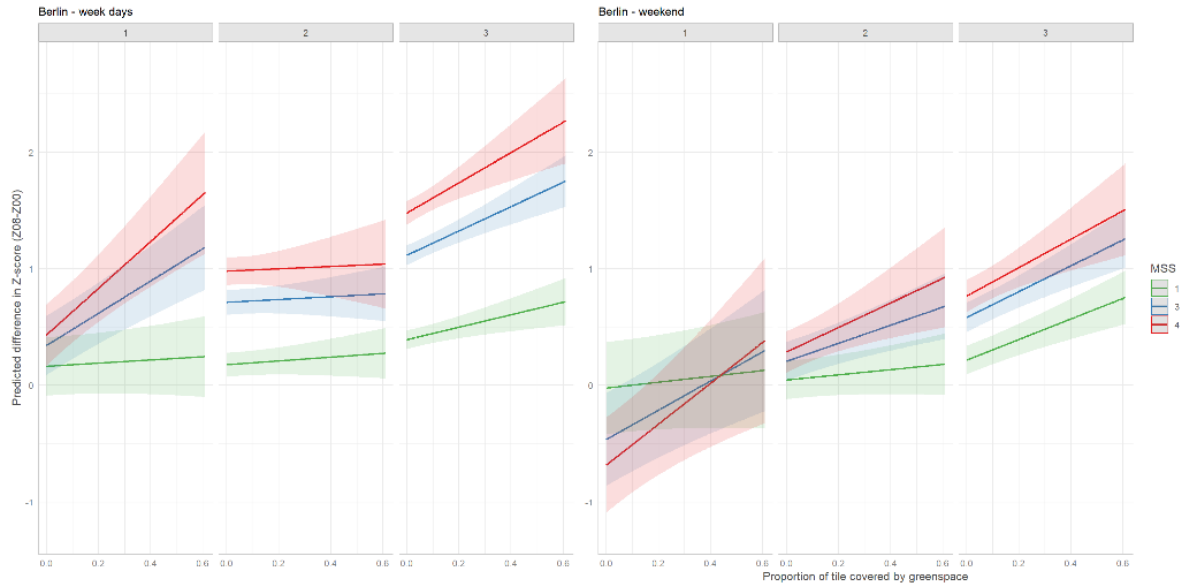

Figure 2. Predicted difference in Z-score ($Z_{08-00}$) in London, Paris, and Berlin when people were required to stay home (with exceptions) in relation to greenspace coverage and the cities' indices of multiple deprivation (red: deprived, blue: average, green: affluent) over waves of restrictions (panel 1, 2 and 3) depending on whether the day is a weekend or not. Error bands are 95% confidence intervals. See Supplementary Tables 3-5 for model selection.

**Urban greenspace discourse**

If this tile-level association between changes in population density and greenspace emerged from behavioural changes, we would expect that people would express their motivation to seek greenspace in a manner associated with this enacted behaviour. We therefore sampled the greenspace discourse on social media. People searched more for this topic during COVID-19 (outside of lockdowns) than they did in the same weeks over the 18 months before the pandemic started (Fig. 3a) and they searched significantly more for the 'Park' topic using the Google search engine outside of lockdowns during COVID-19 (mobility restriction levels 0 & 1: Fig. 3b). This effect was the same across all three countries (Supplementary Tables 8-9). People predominantly searched to find out whether parks were opened (top queries associated with the Topic in those periods). Qualitatively, people talked more about enjoying time in parks on Twitter during the lockdowns in all three cities (tweet topics in English, French, and German; Extended Data Figs. 10-12). However, the volume of discourse about parks on Twitter was complex. It depended on the mobility restriction waves in London while it was constant over the pandemic in Berlin and Paris (Fig. 4). The conversation volume significantly decayed with waves in London and was lower during time of lockdowns (Fig. 4). Yet, tweets about parks during lockdowns received on average significantly more likes in all three cities (Extended Data Fig. 13). As we focussed here on geolocated tweets, a small fraction of all tweets, and the proportion of the French and German population using Twitter is substantially smaller than the British Twitter population, inferences from the results for Berlin and Paris are much less certain.

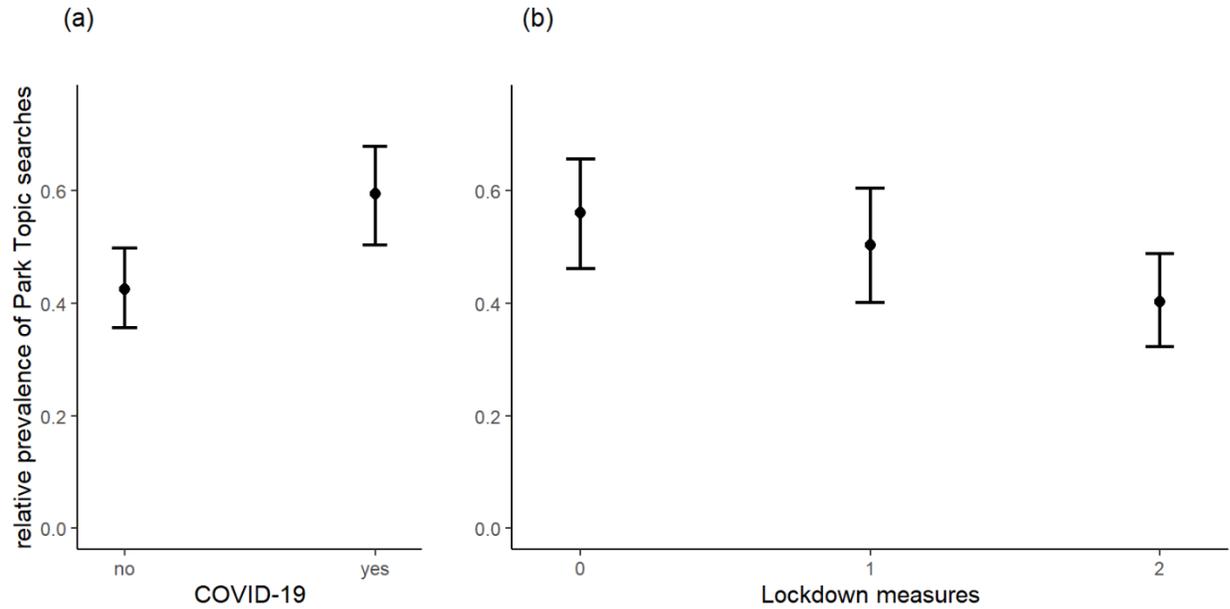

Figure 3. Predicted weekly Google Trends in the 'Park' Topic contrasting the (a) COVID-19 period (Mar 2020 – Aug 2021) to the control period (Aug 2018 – Mar 2020) (Supplementary Table 8) and (b) the periods exposed to different mobility restriction measures (Supplementary Table 9). Searches were significantly more common during COVID-19 and within COVID-19 when people were not confined to their homes. Predictions from generalized mixed effect model, assuming beta distributed errors and 1-lag autocorrelation between weeks among cities.

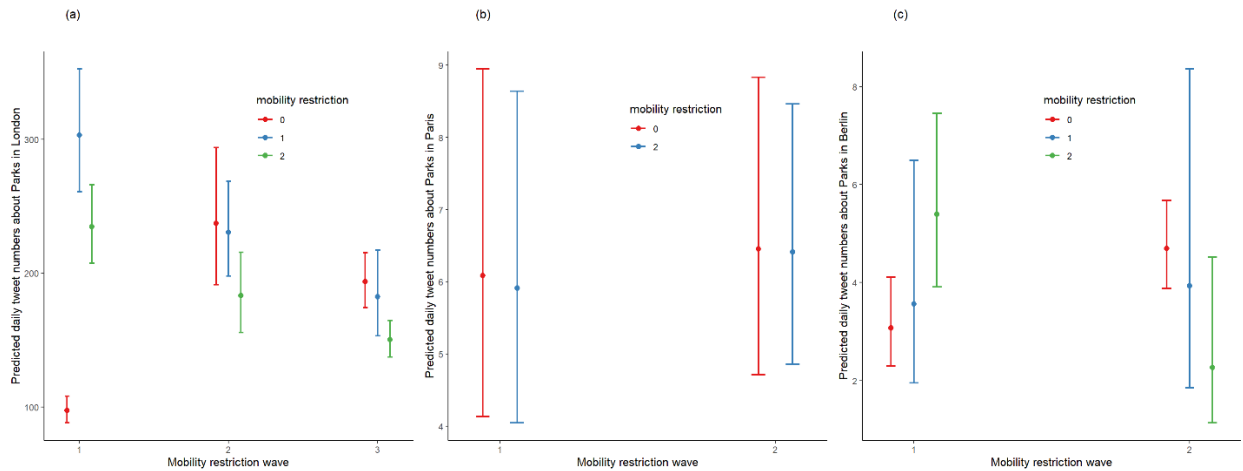

Figure 4. Predicted daily number of tweets about Parks in (a) London, (b) Paris, and (c) Berlin based on a generalized mixed effects model, for each city, assuming a negative binomial distribution of the tweet count residuals and a lag-1 autocorrelation structure between days. There is no support for an effect of lockdown condition and lockdown wave in Paris and Berlin (best models is a constant average number of tweets, Supplementary Table 10), conversely to London (restriction: $\chi^2_2$ = 60.2, p < 0.00001; wave: $\chi^2_2$

= 6.0, p = 0.05; restriction x wave: $\chi^2_4$ = 153.0, p < 0.00001). Mobility restriction: 0: "no measures", 1: "recommend not leaving house", 2: "require not leaving house with exceptions".

Not all greenspace was used in the same manner during the first 20 months of the pandemic. People sought greenspace, searched online for access and commented positively on its use, particularly after lockdowns. As we progressed through the three infection waves, and their associated mobility restrictions, people spent less time in the three cities overall, a general movement pattern detected in other studies[39]. Despite this decline in density, greenspace prevalence at a location explained why that location retained more people during the day. Those that stayed spent more time in locations that had more greenspace during the day ($Z_{08-00}$). However, how they achieved that depended on the level of deprivation of the neighbourhood. In London, only affluent neighbourhoods saw an increase in use with greenspace, while it decreased in deprived neighbourhoods. The difference between neighbourhoods decreased during the weekend, yet overall affluent areas retained, at all times, a greenspace advantage. The greenspace advantage was also observed in Paris, however it disappeared during the weekend. This effect was reversed in Berlin where during the first two waves it appears that people density in affluent neighbourhoods did not change during the day compared to the nights; i.e., people stayed at home, but greenspace preference was observed in more deprived neighbourhoods. Indeed, larger forested parks in the district were also more visited by people from more deprived areas. As we contrast London, Paris, and Berlin it is worth stressing that measures deemed to capture multiple deprivation in Berlin do differ from the other two cities; with a stronger emphasis on who the inhabitants are, particularly whether they are immigrants, rather than infrastructure availability. While immigration status may capture some measures of present socioeconomic status (SES), it masks the richness of SES experience of individuals prior to their arrival in Berlin which would be key in shaping greenspace use[40].

While greenspace availability may be equitable across levels of deprivation by design, the ecosystem services those areas can provide can still be unequal[41]. Affordance plays a key role in this. The same infrastructure may not be able to deliver the same health benefits because needs differ[42,43] or because there is a mismatch between availability and demand[44]. Greenspace access can decrease stress in deprived areas[45]; a mediator for many of the non-communicable diseases that are more prevalent in these urban locations[46]. Yet, to date, there is no information available on how to best design greenspace depending on neighbourhood characteristics to maximise health benefits. As the pandemic continues, and as some countries implement greenspace exposure as health interventions[47,48], we must pay more attention to the heterogeneity of urban greenspace use associated with deprivation. Greenspace design must strive to increase affordance in deprived areas and finally greenspace access, a clear sought and crucial urban infrastructure, must be reported as a dimension of deprivation to better plan for a sustainable urban life.

**Methods**

*Population change sampling*

Facebook has more than 2 billion daily users and provides the broadest social media coverage of individuals across socioeconomic and demographic dimensions ensuring we can capture a more representative sample of people living in the sampled cities. Facebook population aggregated and georeferenced data was made available through an academic license agreement with the Facebook Data for Good programme[33] (https://dataforgood.fb.com/tools/disease-prevention-maps/). This data is available upon request for research purposes through the same procedure from the programme. Data was available for Berlin (2 Apr 2020 to 31 Aug 2021), Paris (26 Apr 2020 to 31 Aug 021) and London (4 Apr 2020 to 31 Aug 2021).

*Public data covariates*

Indices of multiple deprivation (IMDs) had different definition in each city, but these differences captured the variability in the national macro socioeconomic notions of deprivation. All indices considered dimensions of access, demography and income but with measures deemed relevant to each country. None considered greenspace access. We therefore carried out analyses on each city separately, but with a comprehension that changes in IMD could be relatively interpreted in a similar manner in each city (a deprived area in Paris may look different from a deprived area in London, but both can be considered disadvantaged). In all three cities, a tile's proportion of greenspace and IMD did not co-vary (correlation, ρ: London: -0.08, Paris: -0.18 , Berlin: 0.09). Greenspace data was available from open public data[49–51] when those were included, we removed cemeteries from the greenspace spatial data. For indices of multiple deprivation we used IMD at the LSOA level for London[52], FDep for Paris at the IRIS level[53] as the first eigenvector of a weighted principal component analysis of the data available in following standard procedure[54], and the Status index component of MSS (PLR level) for Berlin[55]. The latter is a 4-level categorical variable, while the two former ones are continuous.

*Facebook population density model*

We developed general linear mixed effects models accounting for the repeated daily sampling of tiles as crossed random effects of tiles and date and assuming a Gaussian error structure for $Z_{08}$ and $Z_{08\cdot00}$ response variables. Those were fitted using lme4[56] in R. We then engaged in model selection (using $AIC_c$) where we challenged alternative hypotheses that lockdown effects depend on other factors with data. After model selection and validation, effects were interpreted from tables of contrasts and visualisation of model predictions.

We also developed two ancillary model sets. The first one assessed whether greenspace-rich tiles (main urban parks and gardens) were more visited during the day ($Z_{08}$) when people were required to stay at home. Greenspace-rich tiles were defined as tiles for which the proportion of greenspace cover was greater than the 90% quantile of the distribution of this variable in each respective city (Figure S2b). The second set assessed from where people visiting forests in the Berlin district came. To do so, we used the Facebook movement data[39] which quantifies changes in movement between tiles compared to pre-pandemic movement rate (Z-score). We identified forested tiles in the Berlin district[57], which do not qualify as urban public greenspace and are not categorised as such in the Berlin greenspace data. And determine the proportion of tiles covered in forests. We then identified the MSS status of tiles from which people visiting forested tiles came. We finally assessed whether the movement $Z_{08}$ depended on the MSS of the starting tile and the proportion of forest at the arriving tile. Berlin's greenspace and forest data were accessed thanks to available tailored API code[58]

*Text analyses*

We searched Google Trends[59] for the period 30 August 2018 to 30 August 2021 for the Topic "Park". Trends Topics are an efficient shortcut to deploy a multilingual regular expression search for terms associated with a topic often searched. We limited those searches to the highest spatial resolution available for the three cities: England for London, Berlin region for Berlin, and region Île-de-France for Paris. For this timespan, the temporal resolution of relative search volume was a week. The values represented relative weekly search volume for the Park Topic, where the week with the largest volume was assigned by Google Trends a value of 100, and all others are scaled to be between zero (no weekly searches) and 100. Given this value is continuous and bounded, we divided the value by 100 and assumed beta-distributed residuals when modelling its variance. We then fitted generalized mixed effect models to this relative search prevalence assuming an autocorrelation structure in searches, with a lag of 1 week within cities using glmmTMB[60]. We first assessed whether during the COVID-19 period (Mar 2020-Aug 2021) searches differed depending on the lockdown measures in the cities (using the Oxford COVID-19 GRT categorical variable). We then assessed whether search volume differed during the COVID-19 period outside of lockdown periods compared to the control period (Sept 2018 - Mar 2020).

We searched archived tweets geo-located in the three cities using the academic track of the Twitter API for tweets mentioning Parks in the three relevant languages. This severely censored the number of tweets returned, as most users do not allow geolocation of their tweets, but it ensured that the sampled tweets originated from the cities. We then assessed whether the number of tweets posted depended on mobility restrictions and waves of restrictions for each city using a generalised linear model assuming a lag of 1 day autocorrelation structure and a negative binomial error distribution implemented in glmmTMB. Best model was selected using AIC. We then fitted structured topic models (stm)[61] to the text of tweets in each cities, assessing whether topic prevalence depended on mobility restrictions. Before stm model fitting to the data, the text was cleaned and stemmed and emoji and emoticons were translated to text following usual text preparation procedures[61]. Finally, we assessed whether public response to tweets varied with mobility restriction as well. We estimated whether the number of likes tweets received depended on the mobility restriction phase during which they were posted. We fitted generalised linear mixed effect models using glmmTMB assuming a negative binomial error structure and including a random effect of tweet topics (defined using the previous stms) to ensure that the effect of topic attractiveness was discounted.

**Acknowledgements.** This study was funded by Danmarks Frie Forskningsfond. RB was supported by a NERC QUADRAT PhD studentship NE/S007377/1. We thank the Facebook Data for Good Programme, particularly Alex Pompe, for providing access to the COVID-19 Facebook Population density and movement data and fruitful discussions on how to best use it. We would also like to thank Twitter for the development of the Twitter API for Academic Research product.

**Data availability.** All open data is accessible via the APIs cited in the Methods section. Twitter data is accessible via the Twitter API for Academic Research product after registration to this product, we provide reproducible code at https://www.github.com/dlusseau/greenspace to return the same dataset as the one we used here once people have access to this product. The Google Trends data can be downloaded directly from Google Trends. The Facebook data is accessible via the partnership programme of the Facebook Data for Good Programme.

**Author contributions.** DL conceived the study and carried out the analyses with support from RB. DL implemented sampling protocols. DL and RB interpreted the data and model outcomes. DL wrote the manuscript with support from RB.

**Competing interests.** The authors have no competing interests to declare.

**Materials & Correspondence.** Correspondence and material requests should be addressed to DL (davlu@dtu.dk).

## Extended Data Figures

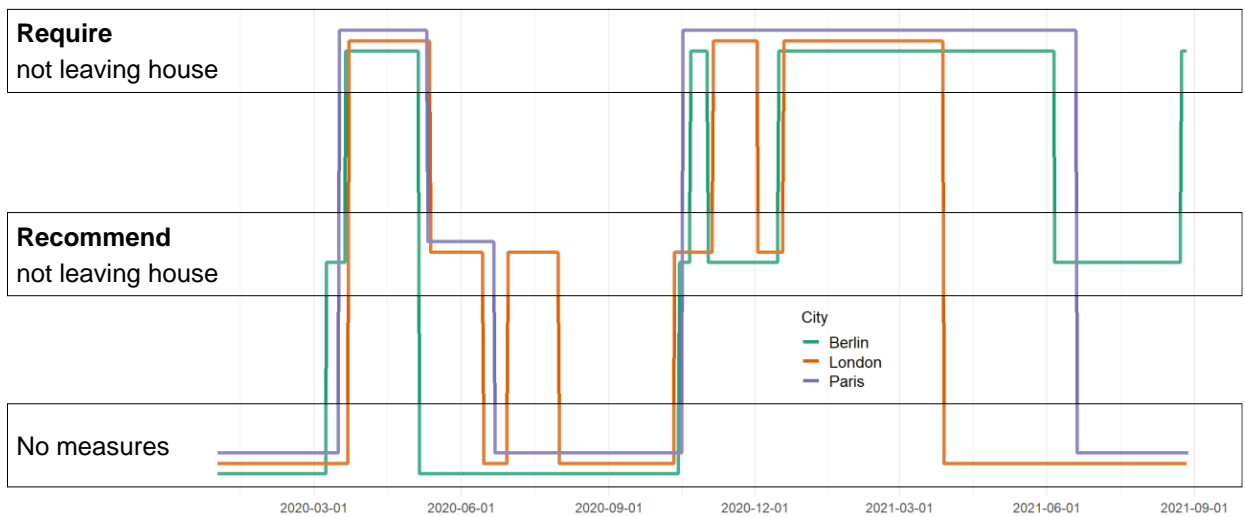

Extended Data Figure 1. Patterns of mobility restrictions in the three cities over the studied period. Data from Oxford COVID-19 Government Response Tracker.

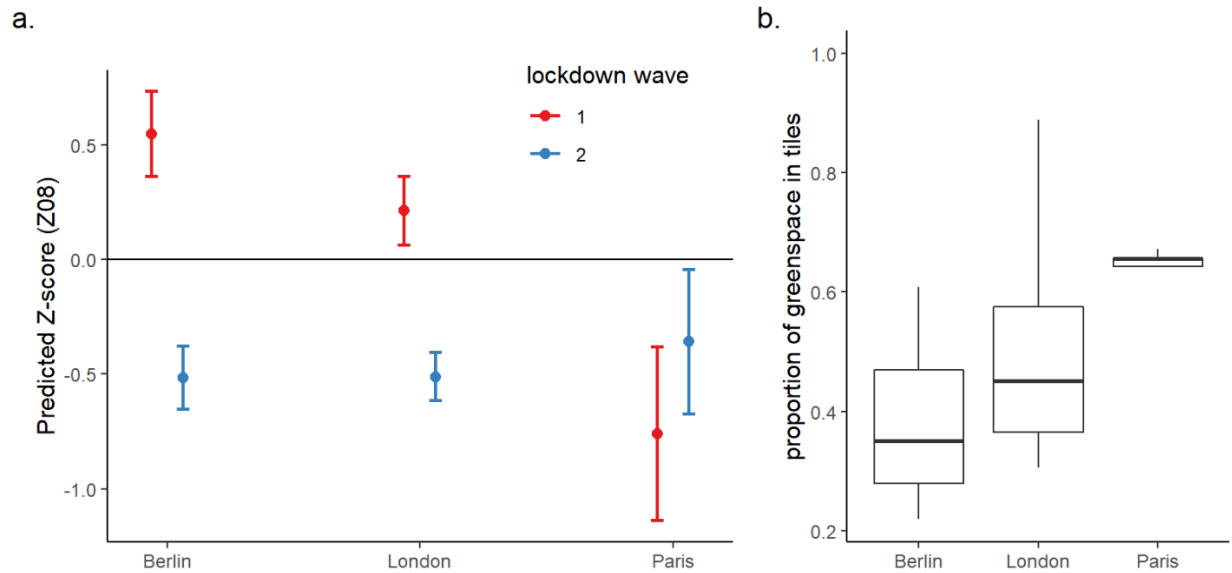

Extended Data Figure 2. a. Predicted change in the Facebook population density during the day ($Z_{08}$) in greenspace rich tiles in the three cities (tiles with proportion of greenspace area in the 90% quantile of the greenspace proportion distribution in tiles in each respective city) during periods when people were required not to leave home (Supplementary Table 2): predicted $Z_{08}$ depending on the lockdown wave (first wave of requirements to not leave home, 1, or subsequent waves, 2, see Extended Data Figure 1). Note public greenspace was closed in Paris during the first wave. Error bars are 95% confidence intervals. b. observed difference in greenspace coverage of greenspace rich tiles in the three cities.

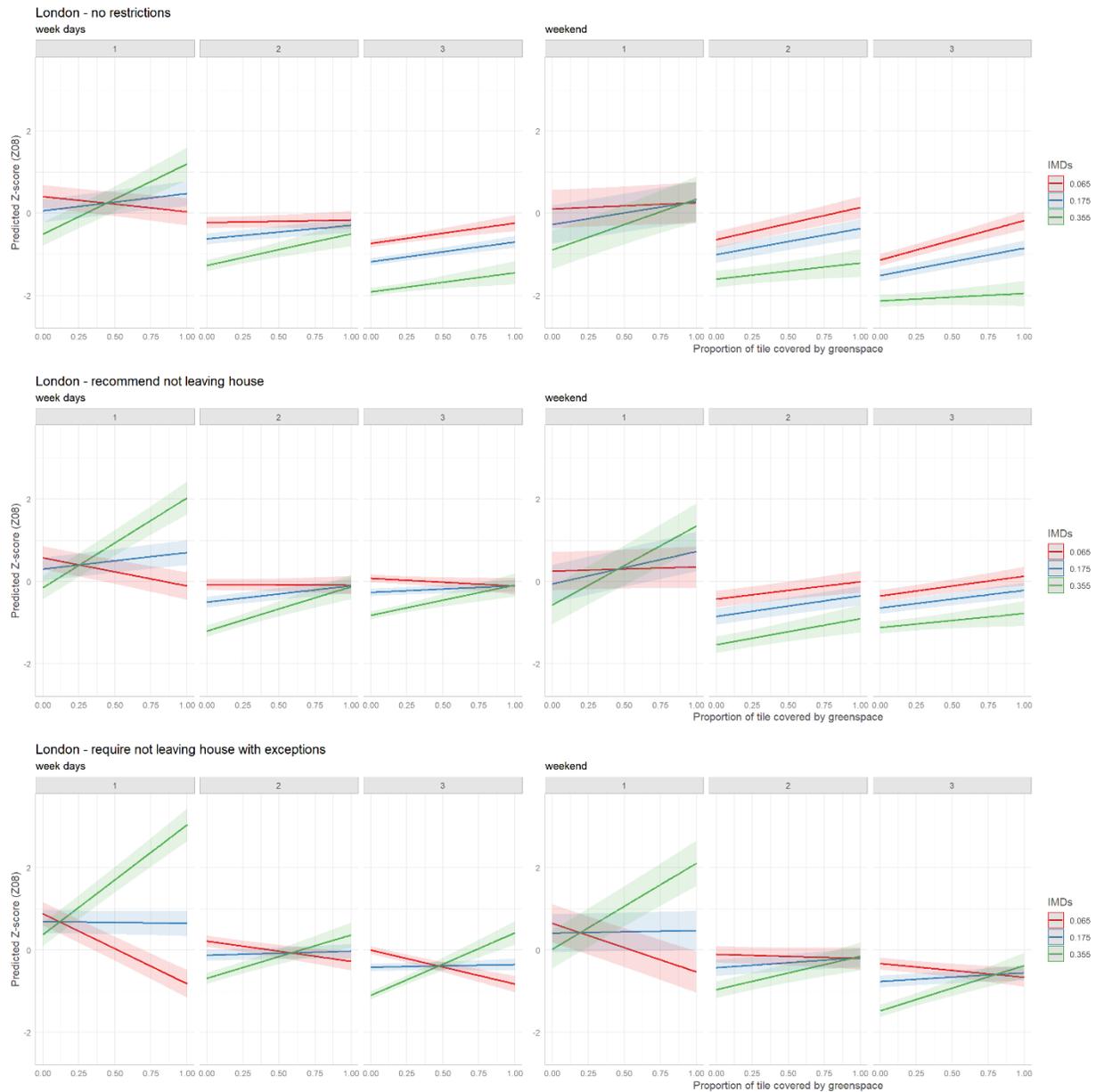

Extended Data Figure 3. Predicted daily $Z_{08}$ score in London in relation to greenspace coverage and the Index of Multiple Deprivation (IMD) over 3 waves of restrictions (panel 1, 2 and 3) depending on whether the day is a weekend or not, and whether it is a lockdown period, mobility restriction is only recommended, or there are no restrictions (Supplementary Table 3, model 16). Error bands are 95% confidence intervals. Predictions from a general linear mixed effect model assuming a Gaussian error distribution with crossed random effects of tile id and date to account for the sampling structure. IMD values are representative of deprived areas (0.065), average areas (0.175) and affluent areas (0.355).

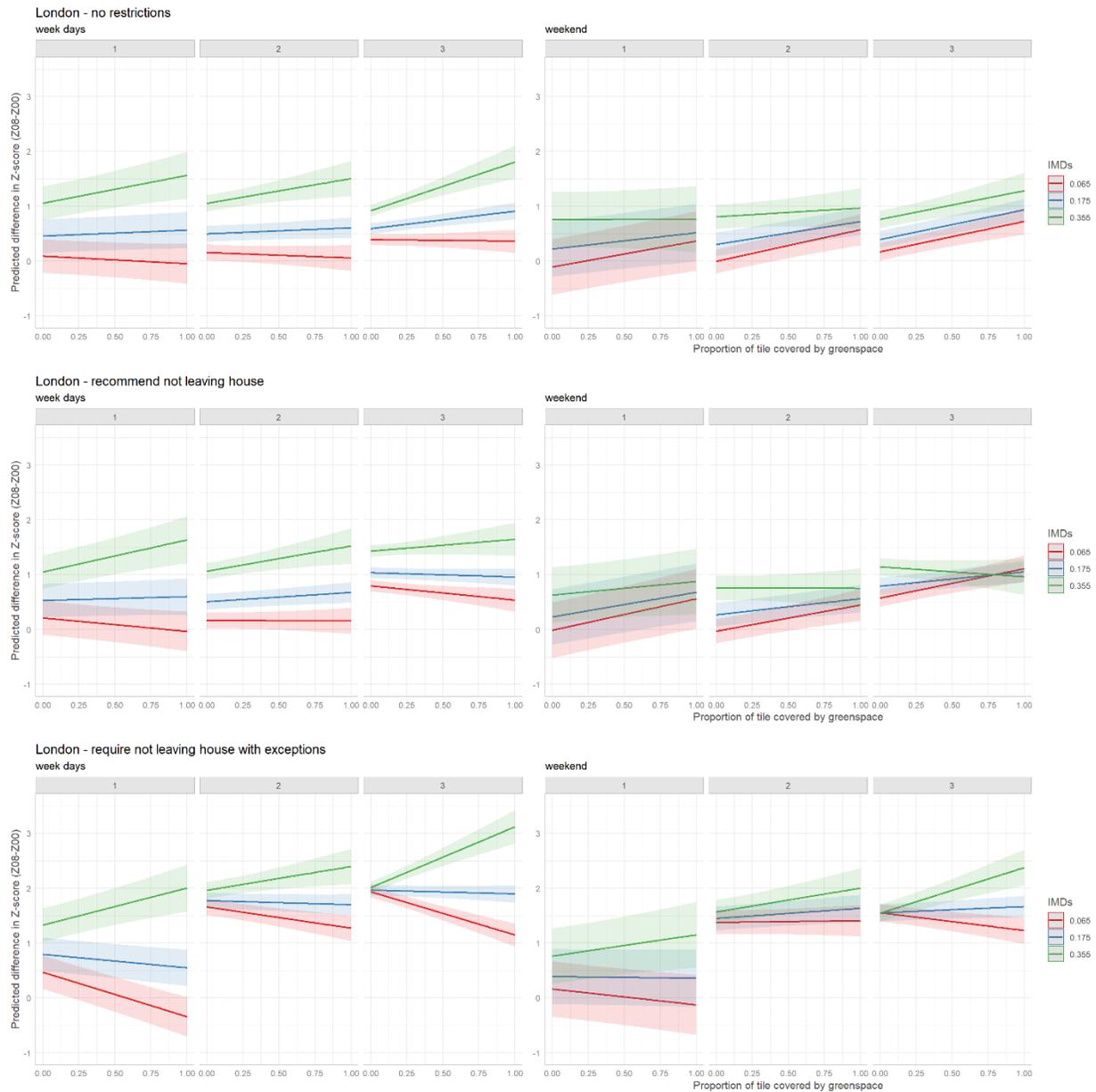

Extended Data Figure 4. Predicted daily difference in Z score ($Z_{08-00}$) in London in relation to greenspace coverage and the Index of Multiple Deprivation (IMD) over 3 waves of restrictions (panel 1, 2 and 3) depending on whether the day is a weekend or not, and whether it is a lockdown period, mobility restriction is only recommended, or there are no restrictions (Supplementary Table 3, model 16). Error bands are 95% confidence intervals. Predictions from a general linear mixed effect model assuming a Gaussian error distribution with crossed random effects of tile id and date to account for the sampling structure. IMD values are representative of deprived areas (0.065), average areas (0.175) and affluent areas (0.355).

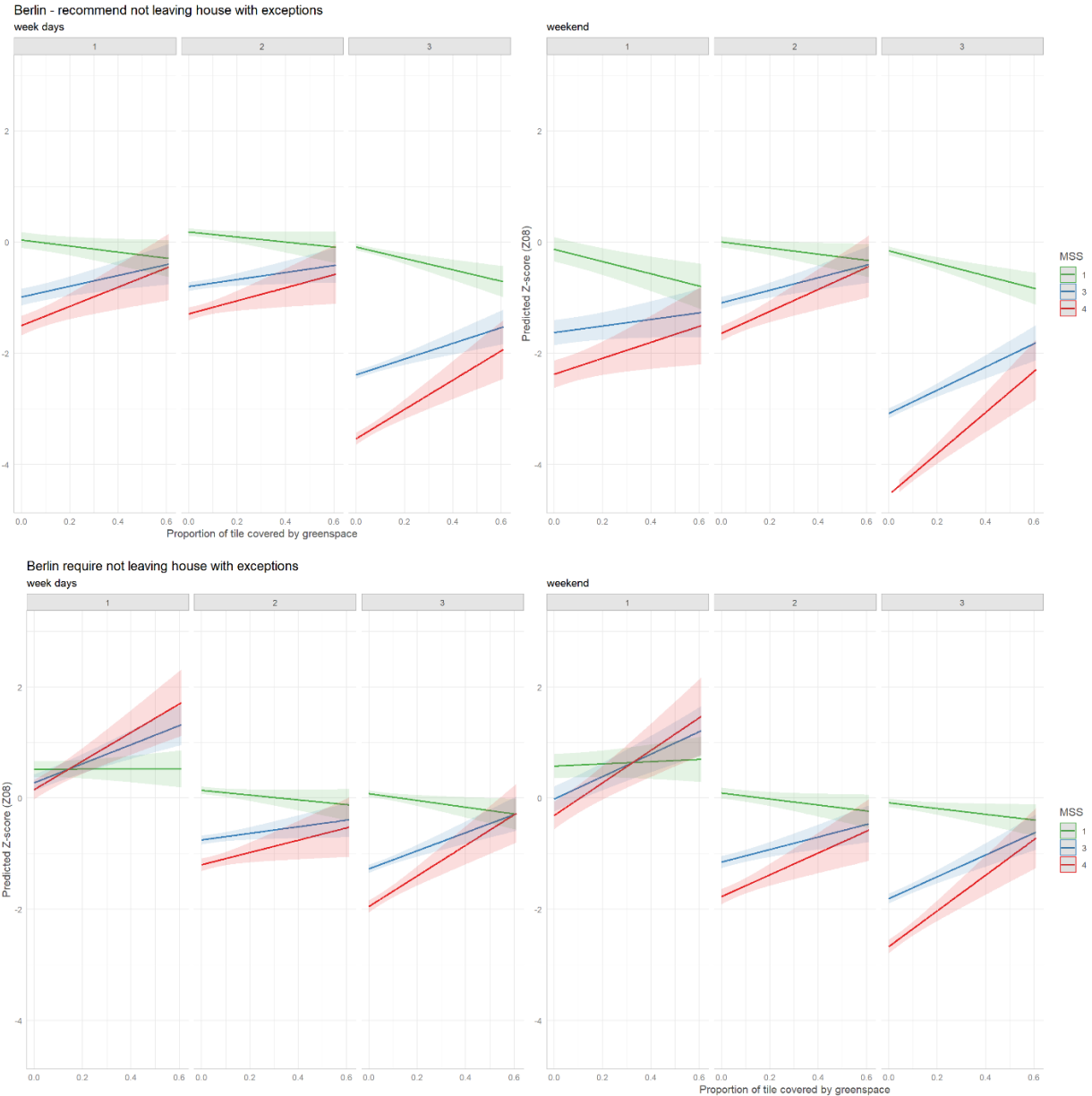

Extended Data Figure 5. Predicted daily $Z_{08}$ score in Berlin in relation to greenspace coverage and the Status Index component of the MSS deprivation index over 3 waves of restrictions (panel 1, 2 and 3) depending on whether the day is a weekend or not, and whether it is a lockdown period or mobility restriction is only recommended (Supplementary Table 4, model 16). Error bands are 95% confidence intervals. Predictions from a general linear mixed effect model assuming a Gaussian error distribution with crossed random effects of tile id and date to account for the sampling structure. MSS categorical values represent deprived areas (4), average areas (3) and affluent areas (1).

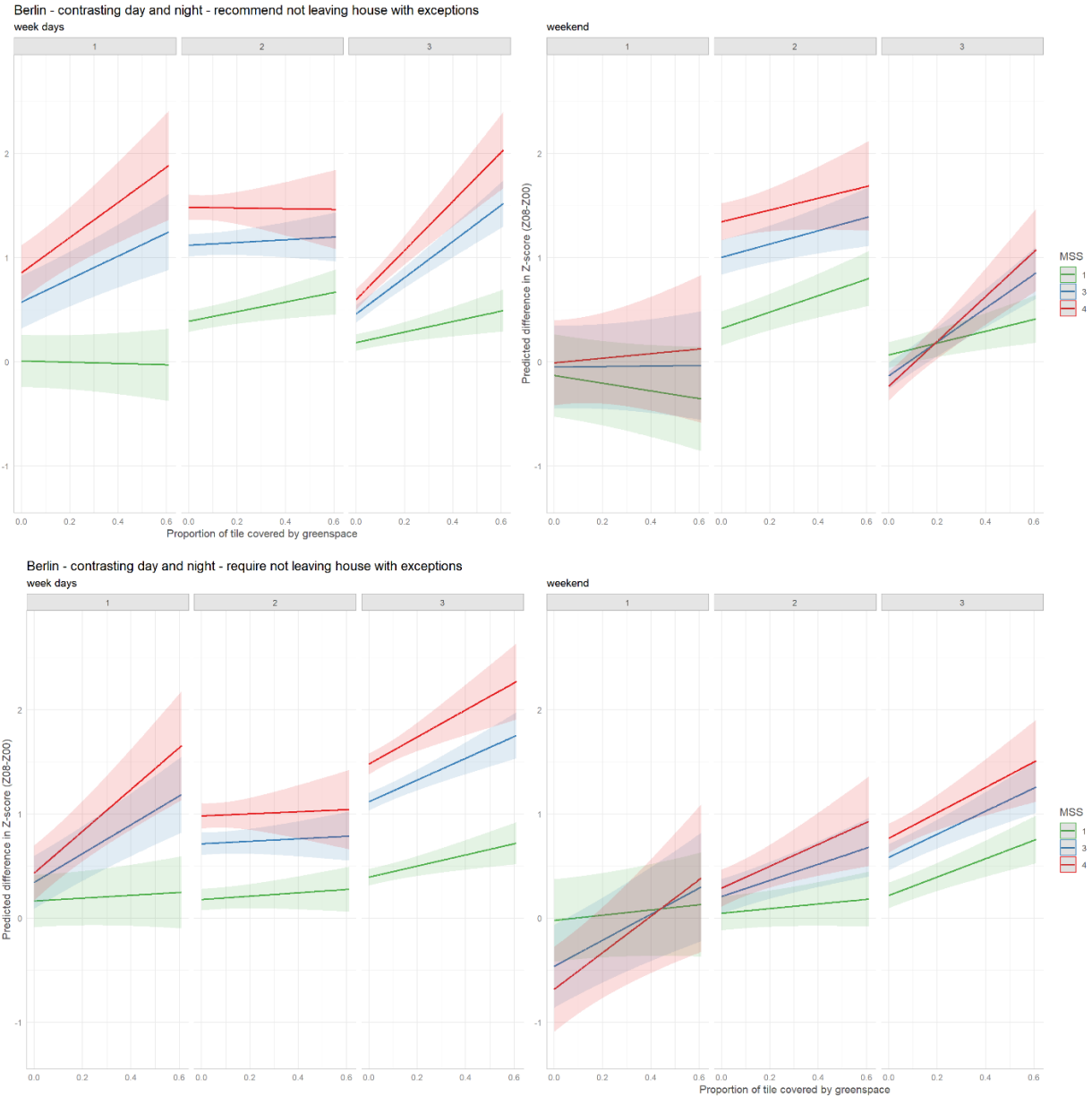

Extended Data Figure 6. Predicted daily difference in Z score ($Z_{08-00}$) in Berlin in relation to greenspace coverage and the Status Index component of the MSS deprivation index over 3 waves of restrictions (panel 1, 2 and 3) depending on whether the day is a weekend or not, and whether it is a lockdown period or mobility restriction is only recommended (Supplementary Table 4, model 16). Error bands are 95% confidence intervals. Predictions from a general linear mixed effect model assuming a Gaussian error distribution with crossed random effects of tile id and date to account for the sampling structure. MSS categorical values represent deprived areas (4), average areas (3) and affluent areas (1).

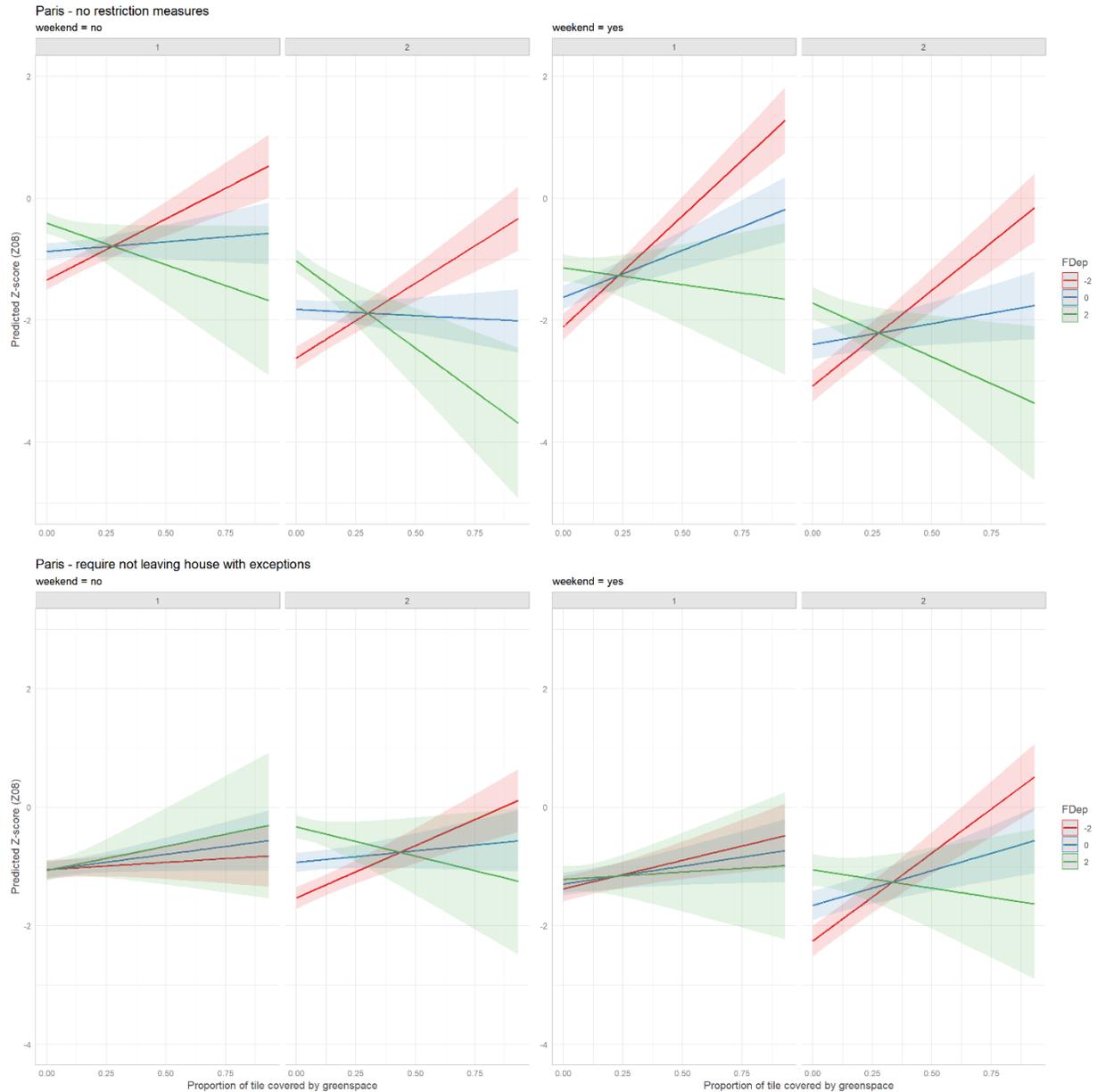

Extended Data Figure 7. Predicted daily $Z_{08}$ score in Paris in relation to greenspace coverage and the French Deprivation Index (FDep) over 2 waves of restrictions (panel 1 and panel 2) depending on whether the day is a weekend or not, and whether it is a lockdown period or mobility is not restricted (Supplementary Table 5, model 16). Error bands are 95% confidence intervals. Predictions from a general linear mixed effect model assuming a Gaussian error distribution with crossed random effects of tile id (n = 644, $\sigma^2$ = 0.695) and date (n = 448, $\sigma^2$ = 0.274, $\sigma^2_{residuals}$ = 0.545) to account for the sampling structure. FDep values are representative of deprived area (-2), average areas (0) and affluent areas (+2).

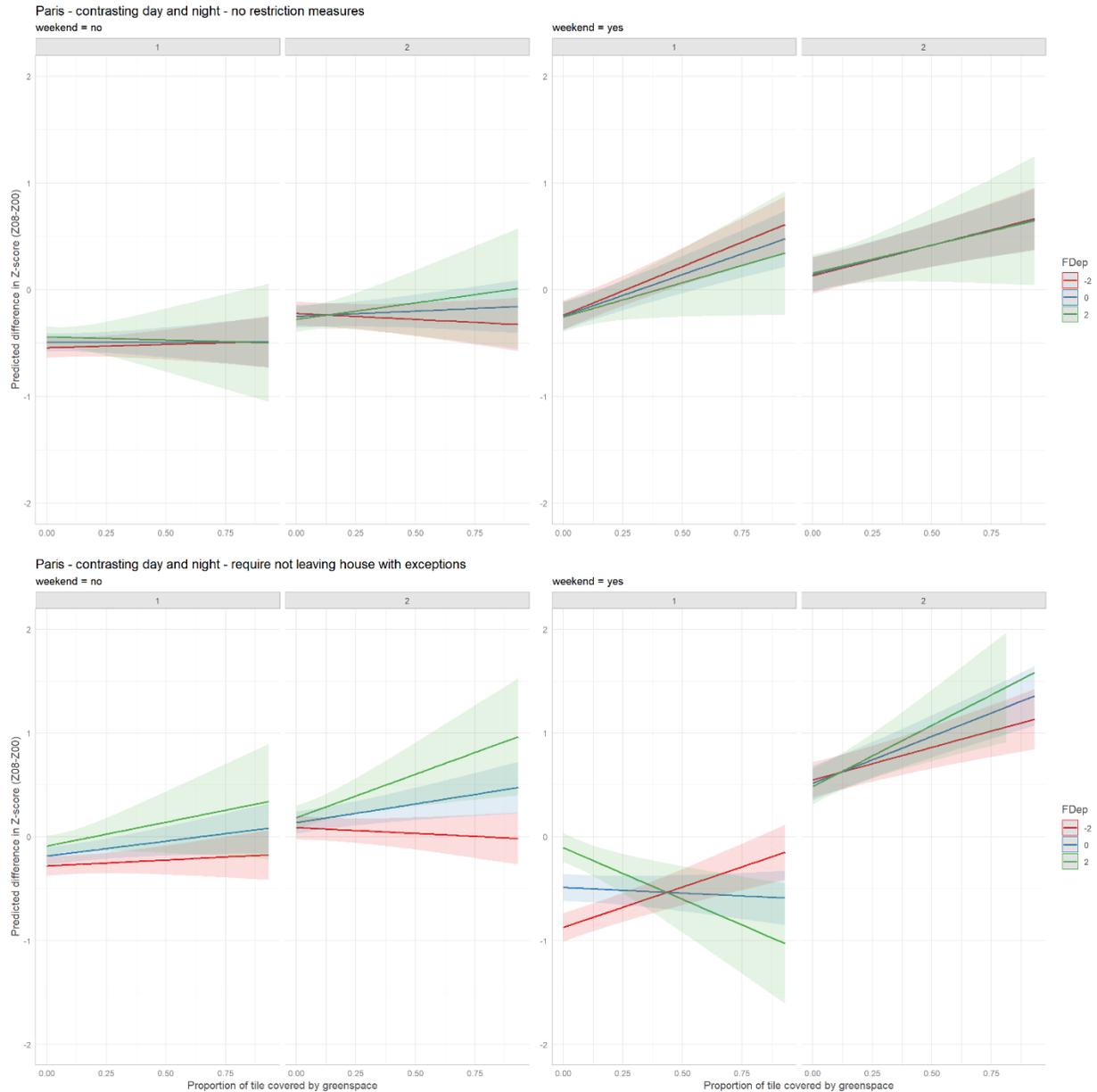

Extended Data Figure 8. Predicted daily difference in Z score ($Z_{08-00}$) in Paris in relation to greenspace coverage and the French Deprivation Index (FDep) over 2 waves of restrictions (panel 1 and panel 2) depending on whether the day is a weekend or not, and whether it is a lockdown period (mobility restriction 2, bottom row) or mobility is not restricted (mobility restriction 0, top row) (Supplementary Table 5, model 16). Error bands are 95% confidence intervals. Predictions from a general linear mixed effect model assuming a Gaussian error distribution with crossed random effects of tile id (n = 644, $\sigma^2$ = 0.136) and date (n = 446, $\sigma^2$ = 0.135, $\sigma^2_{residuals}$ = 0.451) to account for the sampling structure. FDep values are representative of deprived area (-2), average areas (0) and affluent areas (+2).

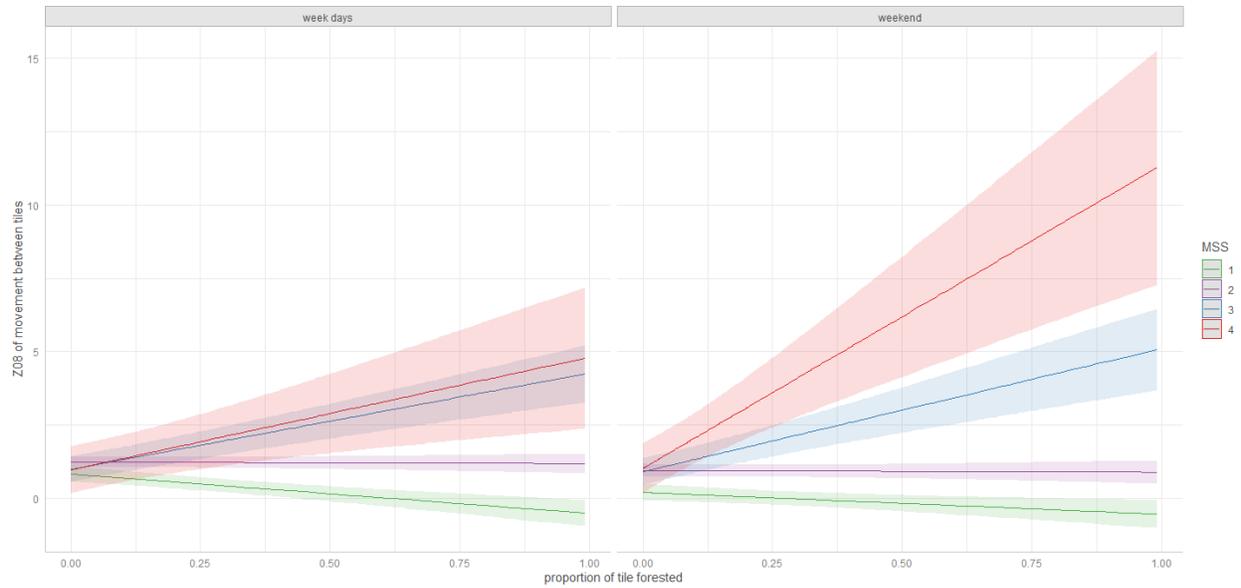

Extended Data Figure 9. Predicted departure from pre-pandemic traffic of people movement in the Berlin district during the day (Z-score, $Z_{08}$) when people were required to stay home from starting tiles with a given MSS to a forested tile with a given $p_{forest}$ depending on whether the movement occurred on a weekend or not (best model, Supplementary Tables 6 & 7). Error bands are 95% confidence intervals. Only tiles with some registered forests ($p_{forest}$>0) are considered.

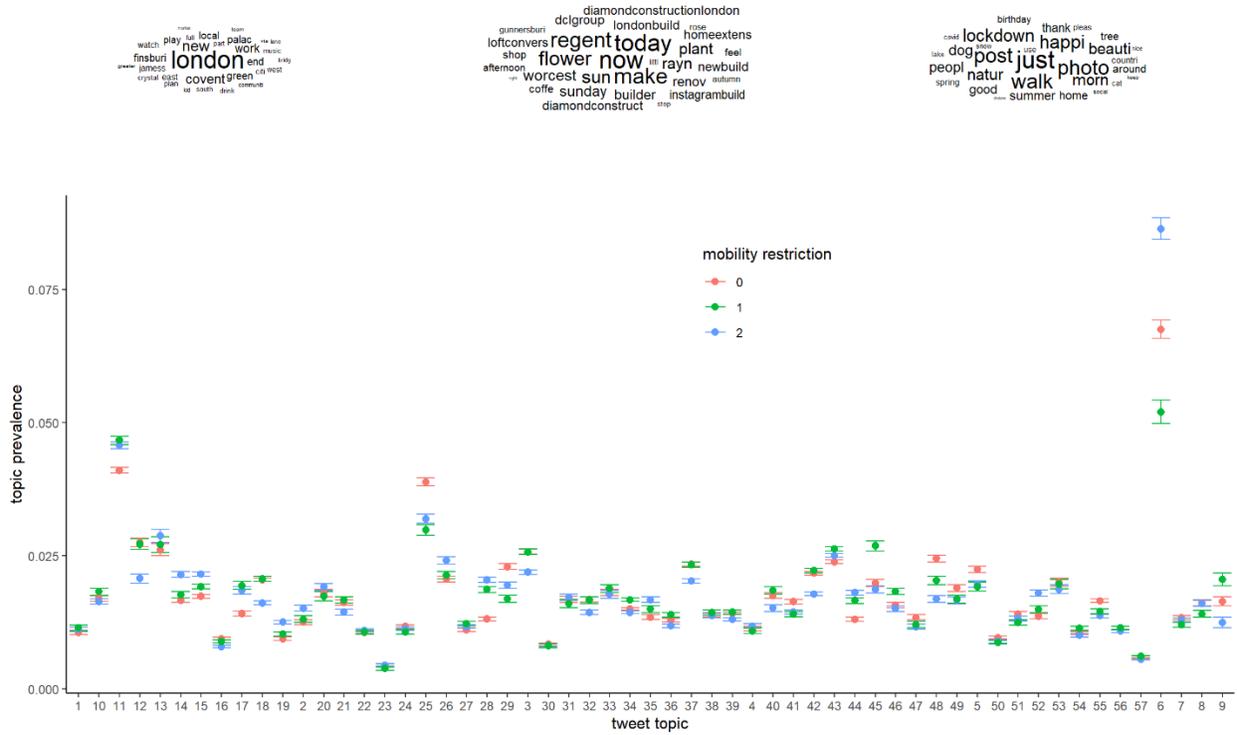

Extended Data Figure 10. Predicted prevalence of topics detected in the structural topic model of English tweets posted in London during the pandemic depending on lockdown conditions. Error bars are 95% confidence intervals. Wordcloud of topics significantly more prevalent in each lockdown condition. Word size is relative to word frequency.

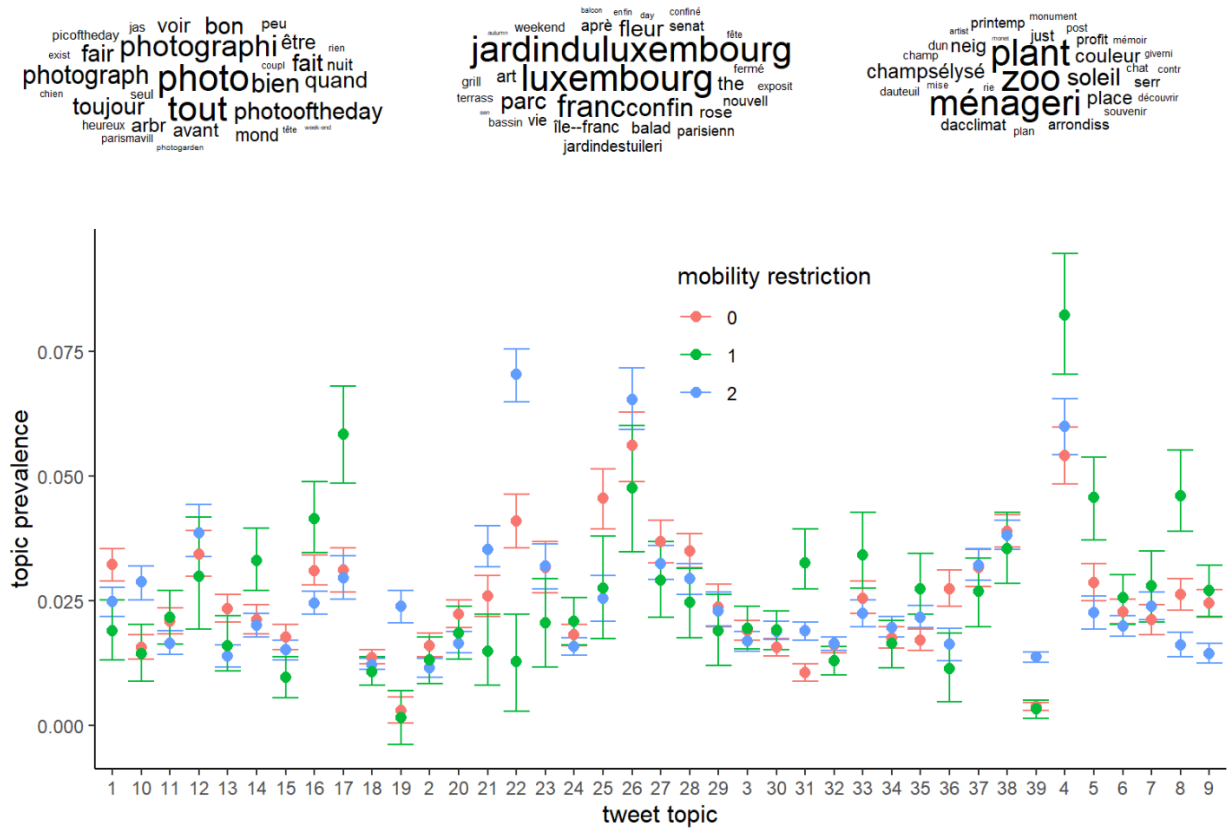

Extended Data Figure 11. Predicted prevalence of topics detected in the structural topic model of French tweets posted in Paris during the pandemic depending on lockdown conditions. Error bars are 95% confidence intervals. Wordcloud of topics significantly more prevalent in each lockdown condition. Word size is relative to word frequency.

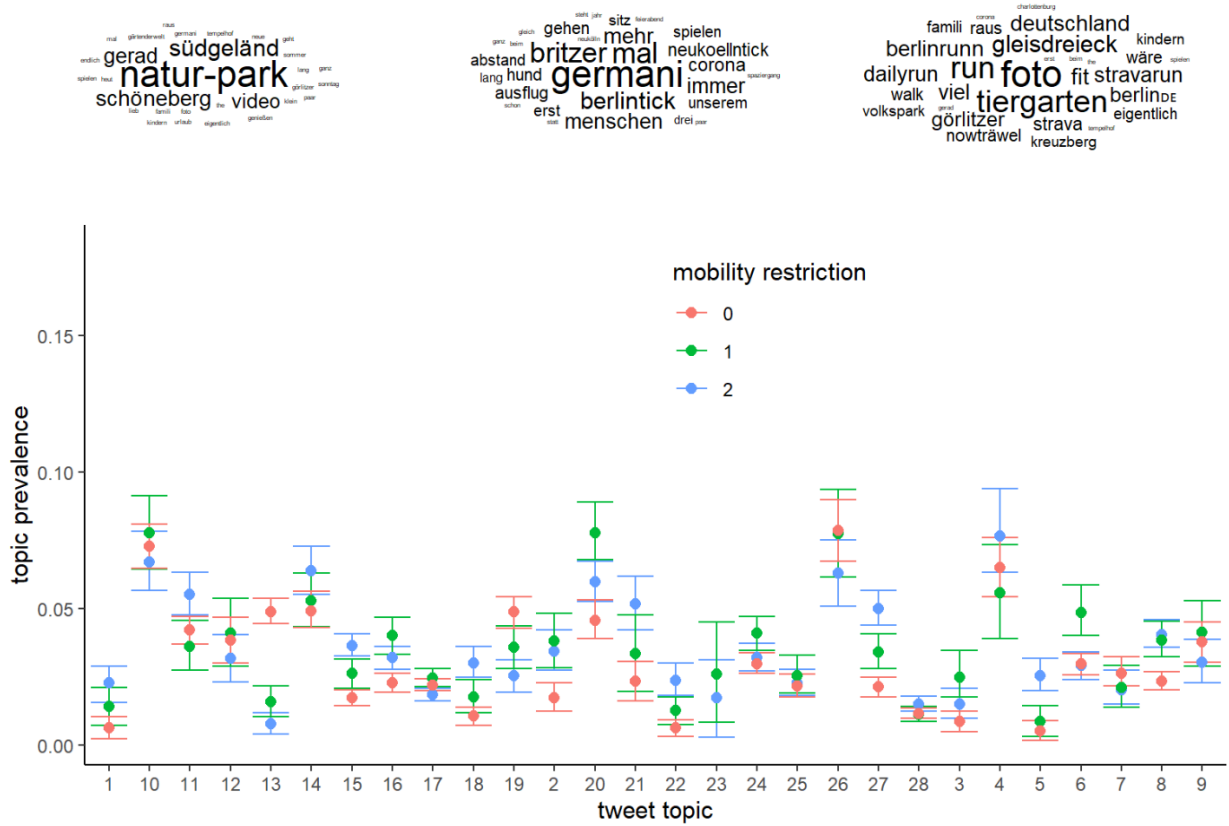

Extended Data Figure 12. Predicted prevalence of topics detected in the structural topic model of German tweets posted in Berlin during the pandemic depending on lockdown conditions. Error bars are 95% confidence intervals. Wordcloud of topics significantly more prevalent in each lockdown condition. Word size is relative to word frequency.

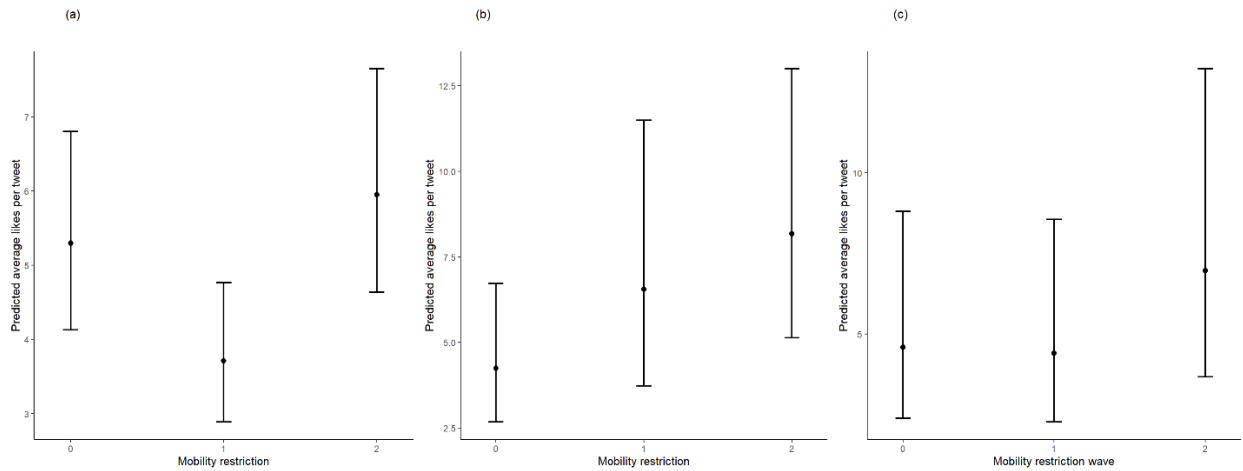

Extended Data Figure 13. Predicted average number of likes a tweet has received depending on the mobility restriction phase for (a) London (restriction: $\chi^2_2$ = 458.4, p < 0.00001), (b) Paris (restriction: $\chi^2_2$ = 29.1, p < 0.00001), and (c) Berlin (restriction: $\chi^2_2$ = 9.77, p= 0.008). Partial effect of mobility restriction phase accounting for a random effect of the Topic of the tweet using a generalized mixed-effect model assuming negative binomial distributed errors. Models include a random effect of the topic discussed in the tweet (Extended Data Figure 10-S12; London: n = 56 topics, $\sigma^2$ =0.895; Paris: n = 36 topics, $\sigma^2$ = 1.384; Berlin: n = 23 topics, $\sigma^2$ = 1.947). 0: "no measures", 1: "recommend not leaving house", 2: "require not leaving house with exceptions".

**Supplementary Tables**

Supplementary Table 1. Selection of models based on Akaike Information Criterion (AIC) to determine factors associated with the daytime Z-score ($Z_{08}$) of greenspace-rich tiles in the three cities during lockdowns (when people were required not to leave home) when considering the city, whether the day was a weekend or not, and whether this was the first lockdown or not ('wave', Extended Data Figure 1). Models are general linear mixed effects models with a random effect of date in city and the id of the tile in the city.

| model | Fixed terms | df | AIC | ΔAIC |
|-------|-------------|-----|-----------|-------|
| 1 | Weekend | 5 | 735986.0 | 219.9 |
| 2 | Weekend x city | 9 | 736000.0 | 233.9 |
| 3 | Wave x city | 9 | **735766.1** | **0** |
| 4 | Wave x city + weekend | 10 | 735772.6 | 6.5 |
| 5 | Wave x city x weekend | 15 | 735785.7 | 19.6 |

Supplementary Table 2. Analysis of Deviance table for the best model describing the variance in $Z_{08}$ for greenspace rich tiles during lockdowns (Supplementary Table 1 model 3). 237,261 observations, variance associated with date in city (0.17), tile in city (1.34), and residuals (1.25).

| terms | $\chi^2$ | df | p-value |
|-------|----------|-----|-------------|
| Wave | 179.0 | 1 | <0.000001 |
| City | 0.38 | 2 | 0.82 |
| Wave x city | 115.5 | 2 | <0.000001 |

Supplementary Table 3. Selection of models based on Akaike Information Criterion (AIC) to determine factors associated with the variance in Facebook population density in tiles in London during the day ($Z_{08}$) and contrasting day and night densities ($Z_{08-00}$). General linear mixed effect models assuming a gaussian residual distribution and a random effect of date (day) and tile ID (quadkey code). Fixed terms are restriction (restriction categorical level 0, 1 and 2, Extended Data Figure 1), wave (the categorical sequential number of the restriction level being repeated, Extended Data Figure 1), whether the day was a weekend or not, greenspace (the proportion of public greenspace in the tile), and IMD, the British multiple deprivation continuous index.  Models retained are in bold.

| model | Fixed terms | $Z_{08}$ | | $Z_{08-00}$ | |
|---|---|---|---|---|---|
| | | AIC | ΔAIC | AIC | ΔAIC |
| 1 | restriction | 17613404 | 61299.4 | 19283345 | 31239.2 |
| 2 | restriction x wave | 17613182 | 61077.9 | 19283233 | 31127.2 |
| 3 | restriction x weekend | 17613394 | 61289.5 | 19283335 | 31229.5 |
| 4 | restriction x wave x weekend | 17613171 | 61067.0 | 19283237 | 31131.9 |
| 5 | restriction x IMD | 17598882 | 46777.8 | 19274587 | 22482.0 |
| 6 | restriction x wave x IMD | 17577959 | 25854.4 | 19261295 | 9189.4 |
| 7 | restriction x weekend x IMD | 17598259 | 46155.0 | 19272875 | 20769.9 |
| 8 | restriction x wave x weekend x IMD | 17576565 | 24460.7 | 19259049 | 6943.4 |
| 9 | restriction x greenspace | 17607982 | 55877.5 | 19281672 | 29566.9 |
| 10 | restriction x wave x greenspace | 17606865 | 54760.7 | 19280743 | 28637.2 |
| 11 | restriction x weekend x greenspace | 17606990 | 54885.9 | 19280494 | 28388.2 |
| 12 | restriction x wave x weekend x greenspace | 17605781 | 53676.3 | 19279523 | 27417.7 |
| 13 | restriction x IMD x greenspace | 17583367 | 31262.8 | 19270830 | 18724.7 |
| 14 | restriction x wave x IMD x greenspace | 17556214 | 4110.2 | 19256606 | 4500.4 |
| 15 | restriction x weekend x IMD x greenspace | 17580271 | 28167.1 | 19266894 | 14788.4 |
| 16 | restriction x wave x weekend x IMD x greenspace | **17552104** | **0.0** | **19252105** | **0.0** |

Supplementary Table 4. Selection of models based on Akaike Information Criterion (AIC) to determine factors associated with the variance in Facebook population density in tiles in Berlin during the day ($Z_{08}$) and contrasting day and night densities ($Z_{08-00}$). General linear mixed effect models assuming a gaussian residual distribution and a random effect of date (day) and tile ID (quadkey code). Fixed terms are restriction (restriction categorical level 1 and 2, Extended Data Figure 1), wave (the categorical sequential number of the restriction level being repeated, Extended Data Figure 1), whether the day was a weekend or not, greenspace (the proportion of public greenspace in the tile), and MSS, the German multiple deprivation categorical index. Models retained are in bold.

| model | Fixed terms | $Z_{08}$ AIC | $Z_{08}$ ΔAIC | $Z_{08-00}$ AIC | $Z_{08-00}$ ΔAIC |
|---|---|---|---|---|---|
| 1 | restriction | 6813881 | 71073.5 | 7270773 | 12517.1 |
| 2 | restriction x wave | 6813390 | 70582.0 | 7270683 | 12426.6 |
| 3 | restriction x weekend | 6813861 | 71053.8 | 7270735 | 12478.6 |
| 4 | restriction x wave x weekend | 6813247 | 70439.5 | 7270636 | 12379.6 |
| 5 | restriction x MSS | 6795684 | 52876.7 | 7269804 | 11548.1 |
| 6 | restriction x wave x MSS | 6749618 | 6810.5 | 7263434 | 5177.5 |
| 7 | restriction x weekend x MSS | 6790895 | 48087.6 | 7266039 | 7783.2 |
| 8 | restriction x wave x weekend x MSS | 6743996 | 1188.8 | 7259113 | 857.0 |
| 9 | restriction x greenspace | 6812785 | 69977.6 | 7270691 | 12435.0 |
| 10 | restriction x wave x greenspace | 6811523 | 68715.8 | 7270189 | 11932.7 |
| 11 | restriction x weekend x greenspace | 6812764 | 69956.6 | 7270663 | 12407.2 |
| 12 | restriction x wave x weekend x greenspace | 6811368 | 68560.5 | 7270070 | 11813.3 |
| 13 | restriction x MSS x greenspace | 6795198 | 52390.9 | 7269748 | 11492.3 |
| 14 | restriction x wave x MSS x greenspace | 6748576 | 5768.7 | 7262655 | 4398.7 |
| 15 | restriction x weekend x MSS x greenspace | 6790269 | 47461.9 | 7265953 | 7696.9 |
| 16 | restriction x wave x weekend x MSS x greenspace | **6742808** | **0.0** | **7258256** | **0.0** |

Supplementary Table 5. Selection of models based on Akaike Information Criterion (AIC) to determine factors associated with the variance in Facebook population density in tiles in Paris during the day ($Z_{08}$) and contrasting day and night densities ($Z_{08-00}$). General linear mixed effect models assuming a gaussian residual distribution and a random effect of date (day) and tile ID (quadkey code). Fixed terms are restriction (restriction categorical level 0,1,2, Extended Data Figure 1), wave (the categorical sequential number of the restriction level being repeated, Extended Data Figure 1), whether the day was a weekend or not, greenspace (the proportion of public greenspace in the tile), and FDEP, the French multiple deprivation continuous index. Models retained are in bold.

| | | $Z_{08}$ | | $Z_{08-00}$ | |
|---|---|---|---|---|---|
| model | Fixed terms | AIC | ΔAIC | AIC | ΔAIC |
| 1 | restriction | 661945.0 | 11354.4 | 596349.7 | 4463.2 |
| 2 | restriction x wave | 660355.8 | 9765.3 | 595145.3 | 3258.8 |
| 3 | restriction x weekend | 661875.5 | 11285.0 | 596282.9 | 4396.4 |
| 4 | restriction x wave x weekend | 660266.6 | 9676.1 | 595060.7 | 3174.1 |
| 5 | restriction x FDep | 661836.1 | 11245.6 | 596251.0 | 4364.5 |
| 6 | restriction x wave x FDep | 656308.3 | 5717.8 | 594822.4 | 2935.9 |
| 7 | restriction x weekend x FDep | 661407.1 | 10816.6 | 595713.7 | 3827.2 |
| 8 | restriction x wave x weekend x FDep | 655848.4 | 5257.9 | 593895.6 | 2009.1 |
| 9 | restriction x greenspace | 661803.3 | 11212.8 | 596335.5 | 4449.0 |
| 10 | restriction x wave x greenspace | 659819.3 | 9228.8 | 595112.3 | 3225.8 |
| 11 | restriction x weekend x greenspace | 657794.5 | 7204.0 | 594261.0 | 2374.4 |
| 12 | restriction x wave x weekend x greenspace | 655613.1 | 5022.6 | 592889.7 | 1003.2 |
| 13 | restriction x FDep x greenspace | 661362.1 | 10771.6 | 596110.1 | 4223.6 |
| 14 | restriction x wave x FDep x greenspace | 655015.7 | 4425.2 | 594630.4 | 2743.9 |
| 15 | restriction x weekend x FDep x greenspace | 657256.0 | 6665.5 | 593826.4 | 1939.8 |
| 16 | restriction xwave x weekend x FDep x greenspace | **650590.5** | **0.0** | **591886.5** | **0.0** |

Supplementary Table 6. Selection of models based on Akaike Information Criterion (AIC) to determine factors associated with the daytime Z-score ($Z_{08}$) for movement to forested tiles during lockdowns (when people were required not to leave home) when considering the multiple deprivation level (MSS) of the tile from which people visiting the forested tiles come ($start_{tile}$), the proportion of the time covered by forests ($p_{forest}$) and whether the day was a weekend or not. Models are general linear mixed effects models with a random effect of date and the id of the starting tile. Only tiles with some registered forests ($p_{forest} > 0$) are considered.

| model | Fixed terms | df | AIC | ΔAIC |
|---|---|---|---|---|
| 1 | $p_{forest}$ + weekend + MSS | 9 | 65631.4 | 159.7 |
| 2 | $p_{forest}$ x weekend x MSS | 19 | **65471.7** | **0** |
| 3 | $p_{forest}$ x weekend | 7 | 65663.7 | 191.9 |
| 4 | $p_{forest}$ | 5 | 65679.0 | 207.3 |
| 5 | $p_{forest}$ x weekend + MSS | 10 | 65635.2 | 163.5 |
| 6 | $p_{forest}$ x weekend + MSS x weekend | 13 | 65597.1 | 125.4 |
| 7 | $p_{forest}$ x weekend + MSS x $p_{forest}$ | 13 | 65519.4 | 47.7 |
| 8 | $p_{forest}$ x weekend + MSS x weekend + MSS x $p_{forest}$ | 16 | 65483.3 | 11.6 |

Supplementary Table 7. Analysis of Deviance table for the best model describing the variance in Z08 of movement to forested tiles (Supplementary Table 6 model 2).

| terms | $\chi^2$ | df | p-value |
|---|---|---|---|
| $p_{forest}$ | 0.640 | 1 | 0.42 |
| MSS | 38.75 | 3 | <0.000001 |
| weekend | 32.02 | 1 | <0.000001 |
| $p_{forest}$ x MSS | 118.60 | 3 | <0.000001 |
| $p_{forest}$ x weekend | 8.80 | 1 | 0.003 |
| MSS x weekend | 51.80 | 3 | <0.000001 |
| $p_{forest}$ x MSS x weekend | 13.64 | 3 | 0.003 |

Supplementary Table 8. Table of contrast for the model predicting the association between Google search volume relative index (Google Trends) and whether searches took place before the pandemic or during the pandemic when people were not required to stay home (Figure 3a). Generalised linear mixed effect model assuming beta-distributed errors and an autocorrelation between weeks within cities with a lag of one week ($\sigma^2 = 0.610$, $\rho = 0.82$; the sampling interval of Google Trends observations). This model (lockdown effect: $\chi^2_1 = 9.20$, p=0.002) was more informative (AIC=-457.6) than one including an effect of the city of origin of the searches (lockdown + city, AIC = -454.4), or one assuming that the effect of lockdown on searches depended on the city (lockdown x city, AIC = -452.1). Model fitted using glmmTMB in R 4.0.3.

| Fixed effect | coefficient | SE | Z | p-value |
|---|---|---|---|---|
| Intercept | -0.30 | 0.150 | -2.01 | 0.04 |
| pandemic – present | 0.68 | 0.225 | 3.03 | 0.002 |

Supplementary Table 9. Table of contrast for the model predicting the association between Google search volume relative index (Google Trends) and lockdown measures during the pandemic (Figure 3b). Generalised linear mixed effect model assuming beta-distributed errors and an autocorrelation between weeks within cities with a lag of one week ($\sigma^2 = 0.670$, $\rho = 0.75$; the sampling interval of Google Trends observations). This model (lockdown effect: $\chi^2_2 = 7.15$, p=0.03) was more informative (AIC=-257.6) than one including an effect of the city of origin of the searches (lockdown + city, AIC = -254.1), or one assuming that the effect of lockdown on searches depended on the city (lockdown x city, AIC = -249.6). Model fitted using glmmTMB in R 4.0.3.

| Fixed effect | coefficient | SE | Z | p-value |
|---|---|---|---|---|
| Intercept | 0.45 | 0.204 | 1.20 | 0.23 |
| Lockdown – level 1 | -0.23 | 0.234 | -1.00 | 0.32 |
| Lockdown – level 2 | -0.64 | 0.248 | -2.58 | 0.009 |

Supplementary Table 10. Selection of models based on Akaike Information Criterion (AIC) to determine factors associated with the daily number of tweets about Parks in London, Paris, and Berlin during the pandemic. Models are generalised linear mixed effect model assuming a negative binomial distribution of daily tweet count residuals and an autocorrelation of tweet count between days with a lag of one day. Model retained for each city in bold.

| Model – fixed terms | Paris | London | Berlin |
|---|---|---|---|
| Constant | **2526.1 (df=4)** | 4591.2 (df=4) | **1895.1 (df=4)** |
| Restriction | 2528.1 (df=5) | 4592.4 (df=6) | 1899.0 (df=6) |
| Wave | 2527.9 (df=5) | 4594.6 (df=6) | 1898.3 (df=6) |
| Restriction + wave | 2529.9 (df=6) | 4596.3 (df=8) | 1902.1 (df=8) |
| Restriction x wave | 2531.9 (df=7) | **4553.3 (df=12)** | 1898.6 (df=12) |